\documentclass [a4paper,10pt] {article}
\usepackage {graphicx}
\usepackage{amsmath}
\usepackage{geometry}
\usepackage{cite}
\usepackage{url}

\begin{document}

\title{A Model Analyzing Life Cycle of Periodical Cicadas}
\author{
Shi Chen\thanks{School of Life Science, Nanjing University,Nanjing
210093 China,chenshinju@gmail.com}, Sheng Bao\thanks{Dept. of
Information Engrg., Nanjing Univ. of P \& T, Nanjing
210003,China,shengbao@ieee.org},Jian-Xiu Chen\thanks{School of Life
Science, Nanjing University, Nanjing210093,China,
chenjx@nju.edu.cn}}
\date{}
 \maketitle

\begin{abstract}
This paper focuses on a mathematical model interpreting the prime
number life cycle of periodical cicadas,\textit{ Magicicada spp.} \
. Changed the viewpoint to predators rather than the prey,this model
fits reality very well by utilizing some principles and
assumption.With the definition of the predator income,natural
selection from predators seems to be the main reason for such a long
life cycle.Consequent solution of this model is exactly the fact of
real nature.
\end{abstract}

\section{Introduction}

Periodical cicadas (\textit{Magicicada spp.}) distribute in North
America.Their nymphae feed on juice of young plant root systems,
living underground for 17 or 13 years. So they are also called 17
years cicadas or 13 years cicadas. Smith's research revealed some species of spiders are nature enemies
of periodical cicadas.

As to the long prime number life cycle of North America cicadas,
there has been dispute among academicians. May\cite{May} and
Murray\cite{Murray}believed it was only coincidence and without
much biological meanings. Japanese scientist
Yoshimura\cite{Yoshimura} argued such long period was to prevent
inbreeding. Lloyd and Dybas\cite{Dybas} suggested some parasites
caused this periodical life cycle.

Other scientists, however, hold the belief that prime number life
cycle could avoid killing from predators. Gould\cite{Gould} argued
if life cycle of cicadas was 12 years, it would possibly encounter
predators whose life cycles are 2,3,4,6 or 12 years; but while the
cicadas have prime number life cycle such as 7 or 11 years, only
those predators who have exactly the same life cycle could devor
them. That is to say prime number life cycle tends to minimize
the loss.

This model is terse and clear.But it is lack of material
interpretation of long period cycle and cannot explain differences
between 11 years and 13 years life cycle: why no 11 year cicadas has
been found in North America. Consequently, a more detailed model
with biological meanings is required. Hoppensteadt and
Keller\cite{Hoppensteadt}set up a differential equation but their
results seemed not very prefect.The calculated life cycle may vary
with different parameter values. Goles group \cite{Goles2001}
\cite{Goles2000} also presented some papers but their research were
based on cicadas whose model setting and discussion are
comparatively difficult. So a predator-centered model is achieved
and solves the fascinating long life cycle problem.

\section{Establishing the Model}
Webb\cite{Webb} in his research assumed that the life cycle of the
ancestor of periodical cicadas varied between 2 to 18 years, while
that of the predator varied between 2 to 5 years. To get more
precise conclusion,we assume the life cycle of
predators is between 2 and 9 years.

According to the opinion of Hoppensteadt and
Keller\cite{Hoppensteadt}, if the life cycle of predator equals
that of the prey(cicada),the predator obtains its maximum
satisfactory:its chance of survival and reproduction. We define its income of the predator is "1".

Denote the life cycle of predator and prey as N and n respectively.

If $n<N$, during the entire life of the predator,only in one year it
has the opportunity to catch the prey.In this year,we say the
predator synchronize with the prey and we define the income of
predator as $\frac{1}{N}$ for this condition.

If $n>N$ and $n$ is a multiple of $N$,($n=kN,k=1,2,\dots$), the
($k-1$)-th generation offspring of the predator has the opportunity
to catch the prey and the ($k-1$)-th generation offspring takes
$\frac{1}{2^{N-1}}$ genetic material of the predator.In this case,we
define the income of predator as $\frac {1}{2^{k-1}}$

If $n>N$ but $n$ is not a multiple of $N$,there is a probability of $\frac{1}{N}$ for
the offspring who takes $\frac{1}{2^{[n/N]-1}}$ genetic material of the predator to
synchronize with prey.To this condition,we define the income as
$$
 \frac {1} {N} \frac{1} {2^{[n/N]-1}}
 $$ where the symbol [  ] means floor.

From the assumptions above, the income of the predator is function
of both the predator and the prey life cycles.Denoting the income as $f$,we obtain

\begin{equation}
f(n,N)=
\begin{cases}
1 & n=N \cr
\frac{1}{N}             & n<N \\
\frac{1}{2^{k-1}}       & n=kN,k=1,2,\dots \\
{1} \over {N 2^{[n/N]-1}} & n>N \ \text{and} \ n \neq kN,k=1,2,\dots\\
\end{cases}
\label{f}
\end{equation}

\section{Solution of this model}

According to Eq.\ref{f},we can calculate out the income of predator in different
permutation of life cycle of the prey and the predator.
The result is listed in Table \ref{income}

\begin{table}[!h]
\begin{center}
\caption[m2]{Income of Predators}%
\label{income}
\begin{tabular}{ c| c c c c c c c c c |c}
 \hline & \multicolumn{9}{c|}{Life Cycle of Predators(unit:year)} &\\
\hline  prey's  & 1& 2& 3& 4& 5& 6& 7& 8& 9& Total \\
\hline  1& 1.0000& 0.5000& 0.3333& 0.2500& 0.2000& 0.1667& 0.1429& 0.1250& 0.1111& 2.8290 \\
 2& 0.5000& 1.0000& 0.3333& 0.2500& 0.2000& 0.1667& 0.1429& 0.1250& 0.1111& 2.8290 \\
 3& 0.2500& 0.5000& 1.0000& 0.2500& 0.2000& 0.1667& 0.1429& 0.1250& 0.1111& 2.7456 \\
 4& 0.1250& 0.5000& 0.3333& 1.0000& 0.2000& 0.1667& 0.1429& 0.1250& 0.1111& 2.7040 \\
 5& 0.0625& 0.2500& 0.3333& 0.2500& 1.0000& 0.1667& 0.1429& 0.1250& 0.1111& 2.4415 \\
 6& 0.0313& 0.2500& 0.5000& 0.2500& 0.2000& 1.0000& 0.1429& 0.1250& 0.1111& 2.6102 \\
 7& 0.0156& 0.1250& 0.1667& 0.2500& 0.2000& 0.1667& 1.0000& 0.1250& 0.1111& 2.1601 \\
 8& 0.0078& 0.1250& 0.1667& 0.5000& 0.2000& 0.1667& 0.1429& 1.0000& 0.1111& 2.4201 \\
 9& 0.0039& 0.0625& 0.2500& 0.1250& 0.2000& 0.1667& 0.1429& 0.1250& 1.0000& 2.0759 \\
 10& 0.0020& 0.0625& 0.0833& 0.1250& 0.5000& 0.1667& 0.1429& 0.1250& 0.1111& 1.3184 \\
 11& 0.0010& 0.0313& 0.0833& 0.1250& 0.1000& 0.1667& 0.1429& 0.1250& 0.1111& 0.8862 \\
 12& 0.0005& 0.0313& 0.1250& 0.2500& 0.1000& 0.5000& 0.1429& 0.1250& 0.1111& 1.3857 \\
 13& 0.0002& 0.0156& 0.0417& 0.0625& 0.1000& 0.0833& 0.1429& 0.1250& 0.1111&\textbf{0.6823} \\
 14& 0.0001& 0.0156& 0.0417& 0.0625& 0.1000& 0.0833& 0.5000& 0.1250& 0.1111& 1.0394 \\
 15& 0.0001& 0.0078& 0.0625& 0.0625& 0.2500& 0.0833& 0.0714& 0.1250& 0.1111& 0.7737 \\
 16& 0.0000& 0.0078& 0.0208& 0.1250& 0.0500& 0.0833& 0.0714& 0.5000& 0.1111& 0.9695 \\
 17& 0.0000& 0.0039& 0.0208& 0.0313& 0.0500& 0.0833& 0.0714& 0.0625& 0.1111& \textbf{0.4344} \\
 18& 0.0000& 0.0039& 0.0313& 0.0313& 0.0500& 0.2500& 0.0714& 0.0625& 0.5000& 1.0003 \\
\hline   Total& 2.0000& 3.4922&3.9271& \textbf{4.0000}& 3.9000& 3.8333& 3.5000& 3.3750& 3.2778&  \\
\hline
\end{tabular}
\end{center}
\end{table}

Assuming the life cycle of ancestor of the prey and the predator is
random distributed in 1 to 18 and 1 to 9 respectively, we can
calculate out income of the predator in different life cycle choice
of the prey and the predator by adding value of $f(n,N)$ in rows and
in columns respectively. The result is listed in the column and row
labeled "Total".

\begin{figure}[!h]
\begin{center}
\includegraphics[scale=0.6]{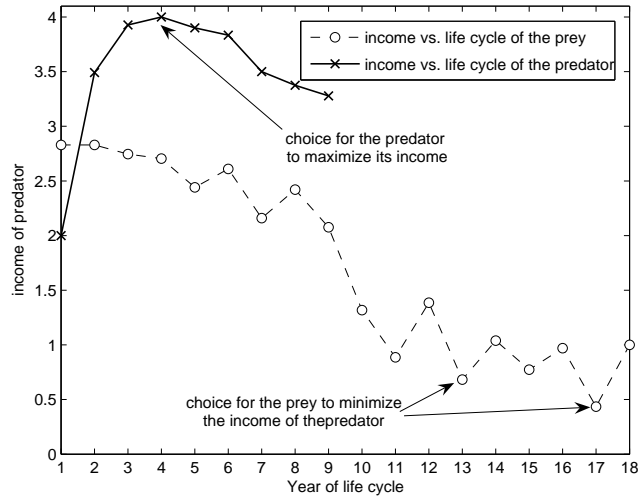}
\caption[17years]{life cycle vs. income}%
\label{vs}
\end{center}
\end{figure}

Data in "Total" column indicates the income of the predator the different life cycle choice
 of the prey whereas data in "Total" row reflects the income of the predator vs. its life
 cycle.They are represented in dash line with round marker and solid line with cross marker
 in Figure \ref{vs} respectively.

According to Figure \ref{vs},the prey choose 13 and 17 years as its life cycle can minimize
the income of predator while the predator choose 4 years as its life cycle can maximize its income.

\section{Accordance with Reality}

From Table \ref{income} and Figure \ref{vs}, the 13 and 17 years
life cycle could reduce the income of predator significantly, which
accords with the fact. Moreover, the troublesome "11" year period problem is
solved that 11 years life cycle is not so efficient in reducing income of predators.

Solid line in Figure \ref{vs} indicates 4 years should be the best choice for predators.Vail\cite{Vail} pointed out one
nature enemy of periodical cicadas \textit{Mud dauber} has life
cycle of 4 years which accords with this model well.

\section{Conclusion}
Unlike traditional researches,this paper is a predator-focused model
to interpret the phenomena of prime number life cycle of periodical
cicadas in north America.The model utilizes some basic principles
and assumptions.Defined the income of predator, this simple fits the
reality very well.Both the fact from \textit{Magicicada spp.} and
its predator supports our conclusion.

\bibliographystyle{IEEEtran}
\bibliography{cicada}

\begin{thebibliography}{10}
\providecommand{\url}[1]{#1}
\csname url@rmstyle\endcsname
\providecommand{\newblock}{\relax}
\providecommand{\bibinfo}[2]{#2}
\providecommand\BIBentrySTDinterwordspacing{\spaceskip=0pt\relax}
\providecommand\BIBentryALTinterwordstretchfactor{4}
\providecommand\BIBentryALTinterwordspacing{\spaceskip=\fontdimen2\font plus
\BIBentryALTinterwordstretchfactor\fontdimen3\font minus
  \fontdimen4\font\relax}
\providecommand\BIBforeignlanguage[2]{{%
\expandafter\ifx\csname l@#1\endcsname\relax
\typeout{** WARNING: IEEEtran.bst: No hyphenation pattern has been}%
\typeout{** loaded for the language `#1'. Using the pattern for}%
\typeout{** the default language instead.}%
\else
\language=\csname l@#1\endcsname
\fi
#2}}

\bibitem{May}
R.~M. May, ``Periodic cicadas,'' \emph{Nature}, vol. 277, pp. 347--349, 1979.

\bibitem{Murray}
J.~D. Murray, \emph{Mathematical Biology}.\hskip 1em plus 0.5em minus
  0.4em\relax Springer-Verlag, 1989.

\bibitem{Yoshimura}
J.~Yoshimura, ``The evolutionary origins of periodical cicadas during ice
  ages,'' \emph{American Naturalist}, vol. 149, pp. 112--124, 1997.

\bibitem{Dybas}
H.~S. Dybas and M.~Lloyd, ``The habitats of 17-year periodical cicadas
  (homoptera: Cicadidae: Magicicada),'' \emph{Ecological Monogr.}, vol.~44, pp.
  279--324, 1974.

\bibitem{Gould}
S.~J. Gould, \emph{Cicadas and the economy of Adam Smith,Ever Since
  Darwin}.\hskip 1em plus 0.5em minus 0.4em\relax Norton Press, 1977.

\bibitem{Hoppensteadt}
F.~C. Hoppensteadt and J.~B. Keller, ``Synchronization of periodical cicada
  emergences,'' \emph{Science}, pp. 335--337, 1976.

\bibitem{Goles2001}
E.~Goles, O.~Schulz, and M.~Markus, ``Prime number selection of cycles in a
  predator-prey model,'' \emph{Complexity}, vol.~6, pp. 33--38, 2001.

\bibitem{Goles2000}
E.~Goles, O.~Schulz, and M.~Markus, ``A biological generator of prime numbers,'' \emph{Nonlinear Phenomena
  in Complex Systems}, pp. 208--213, 2000.

\bibitem{Webb}
G.~F. Webb, ``The prime number periodical cicada problem,'' \emph{Discrete and
  continuous Dynamical Systems , Series B}, vol. 1(3), pp. 387--399, 2001.

\bibitem{Vail}
K.~Vail, ``Mud dauber and cicada killers,'' \emph{Agricultural Extension
  Service of The University of Tennessee}, pp. SP341T--2.5M--9/98(REV)
  E12--2015--00--044--99, 1998.

\end{thebibliography}

\section*{Appendix}
The MATLAB code for calculation and plotting.
\begin{verbatim}
% Matlab code for calculating.Cicada V.2 Last Modified at Oct.9,2005
% Programmed by Sheng Bao  <shengbao@ieee.org> <http://grandlab.cer.net/~bao>
% --This code is distributed under the terms of GNU GPL (General Public License)
% --You can modify it and redistribute it freely but you must keep is open source.

for n=1:18 % prey
    for N=1:9 %predator
        if n==N
            f(n,N)=1;
        elseif n<N
            f(n,N)=1/N;
        elseif (n>N)&&(rem(n,N)==0)
            f(n,N)=2^(1-(n/N));
        else
            f(n,N)=1/(N*2^(fix(n/(N)-1)));
        end
    end
end


for i=1:18
    m1(i)=sum(f(i,:));
end

for i=1:9
    m2(i)=sum(f(:,i));
end

box on plot(m1,'--ko','MarkerEdgeColor','k','MarkerFaceColor',[1 1
1],'MarkerSize',6)

hold on
axis([1 18 0 4.1])
 set(gca,'XTick',1:1:18)
 set(gca,'XTickLabel',{'1','2','3','4','5','6','7','8',...
                       '9','10','11','12','13','14','15','16','17','18'})
plot(m2,'-kx','LineWidth',1,'MarkerEdgeColor','k',...
                            'MarkerFaceColor',[1 1 1],'MarkerSize',8)

legend('income vs. life cycle of the prey',...
        'income vs. life cycle of the predator',3)
xlabel('Year of life cycle')
ylabel('income of predator')
title('')
\end{verbatim}

\end{document}